\documentclass[acmsmall]{acmart}

\usepackage{algorithmic}
\usepackage{graphicx}
\usepackage{textcomp}
\usepackage{xcolor}

\usepackage{makecell}
\usepackage{tabularx}
\usepackage[T1]{fontenc}
\usepackage{comment}
\usepackage{xcolor}
\usepackage{multirow}
\usepackage{array}
\usepackage{soul}
\usepackage{threeparttable}
\usepackage{balance}
\usepackage{diagbox}
\usepackage{ulem}
\usepackage{subfigure}
\usepackage{microtype} 
\usepackage{calc}

\usepackage{enumitem}
\usepackage{url}
\usepackage{xspace}
\usepackage{float}
\usepackage{stfloats}
\usepackage{adjustbox}
\usepackage{amsmath,amssymb,amsfonts}
\usepackage{algorithmic}
\usepackage{graphicx}
\usepackage{textcomp}
\usepackage{xcolor}
\usepackage{multirow}
\usepackage{mdframed}
\usepackage{tcolorbox}
\usepackage{makecell}
\usepackage{threeparttable}
\usepackage{colortbl}
\usepackage{subfigure}
\usepackage{hhline}
\usepackage{pifont}
\usepackage{enumitem}

\usepackage{booktabs}
\usepackage{siunitx}

\usepackage[utf8]{inputenc}
\usepackage[english]{babel}
\usepackage{enumitem} 
\usepackage{bm}

\newcommand\app{\textsc{APIKG4Syn}\xspace}
\newcommand\qwentwo{Qwen2.5-Coder\xspace}
\newcommand\qwenthree{Qwen3\xspace}
\newcommand\deepseek{Deepseek-R1\xspace}

\newcommand\mistral{Mistral\xspace}
\newcommand\gpt{GPT-4o\xspace}
\newcommand\omini{o4-mini\xspace}

\newcommand{\parabf}[1]{\noindent\textbf{#1}}

\newcommand{\Comment}[1]{}

\definecolor{ggray}{HTML}{eff0f0}
\definecolor{gggray}{HTML}{E8E8E8}
\definecolor{ggggray}{HTML}{BEBEBE}

\newcounter{finding}
\newcommand{\finding}[1]{\refstepcounter{finding}
 	\vspace{1mm}
	\begin{mdframed}[linecolor=gray,roundcorner=12pt,backgroundcolor=gray!15,linewidth=3pt,innerleftmargin=2pt, leftmargin=0cm,rightmargin=0cm,topline=false,bottomline=false,rightline = false]
		\textbf{Finding \arabic{finding}:} #1
	\end{mdframed}
	\vspace{1mm}
}

\newcommand{\distance}{5pt}
\AtBeginDocument{%
  }

\setcopyright{cc}
\setcctype{by-nc-nd}
\acmDOI{10.1145/3808133}
\acmYear{2026}
\acmJournal{PACMSE}
\acmVolume{3}
\acmNumber{FSE}
\acmArticle{FSE126}
\acmMonth{7}
\acmSubmissionID{fse26mainb-p376-p}
\received{2025-09-12}
\received[accepted]{2026-03-24}

\begin{document}

\title[KG-Driven Data Synthesis for Low-Resource Software Development: A HarmonyOS Case Study]{Knowledge-Graph-Driven Data Synthesis for Low-Resource Software Development: A HarmonyOS Case Study}

\author{Mingwei Liu}
\orcid{0000-0002-3462-997X}
\affiliation{%
  \institution{Sun Yat-sen University}
  \city{Zhuhai}
  \country{China}
}
\affiliation{%
  \institution{Zhuhai Key Laboratory of Trusted Large Language Models}
  \city{Zhuhai}
  \country{China}
}
\email{liumw26@mail.sysu.edu.cn}

\author{Zheng Pei}
\orcid{0009-0001-5127-7750}
\affiliation{%
  \institution{Sun Yat-sen University}
  \city{Zhuhai}
  \country{China}
}
\affiliation{%
  \institution{Zhuhai Key Laboratory of Trusted Large Language Models}
  \city{Zhuhai}
  \country{China}
}
\email{zhengpei516@gmail.com}

\author{Yanlin Wang}
\authornote{Yanlin Wang is the corresponding author.}
\orcid{0000-0001-7761-7269}
\affiliation{%
  \institution{Sun Yat-sen University}
  \city{Zhuhai}
  \country{China}
}
\affiliation{%
  \institution{Zhuhai Key Laboratory of Trusted Large Language Models}
  \city{Zhuhai}
  \country{China}
}
\email{wangylin36@mail.sysu.edu.cn}

\author{Zihao Wang}
\orcid{0009-0005-2905-9131}
\affiliation{%
  \institution{Sun Yat-sen University}
  \city{Zhuhai}
  \country{China}
}
\affiliation{%
  \institution{Zhuhai Key Laboratory of Trusted Large Language Models}
  \city{Shanghai}
  \country{China}
}
\email{wangzh778@mail2.sysu.edu.cn}

\author{Zikang Li}
\orcid{0009-0003-8024-4995}
\affiliation{%
  \institution{Sun Yat-sen University}
  \city{Zhuhai}
  \country{China}
}
\affiliation{%
  \institution{Zhuhai Key Laboratory of Trusted Large Language Models}
  \city{Zhuhai}
  \country{China}
}
\email{lizk26@mail2.sysu.edu.cn}

\author{Enci Lin}
\orcid{0009-0003-7024-2531}
\affiliation{%
  \institution{Sun Yat-sen University}
  \city{Zhuhai}
  \country{China}
}
\affiliation{%
  \institution{Zhuhai Key Laboratory of Trusted Large Language Models}
  \city{Zhuhai}
  \country{China}
}
\email{linenc@mail2.sysu.edu.cn}

\author{Xin Peng}
\orcid{0000-0003-3376-2581}
\affiliation{%
  \institution{Fudan University}
  \city{Shanghai}
  \country{China}
}
\email{pengxin@fudan.edu.cn}

\author{Zibin Zheng}
\orcid{0000-0002-7878-4330}
\affiliation{%
  \institution{Sun Yat-sen University}
  \city{Zhuhai}
  \country{China}
}
\affiliation{%
  \institution{Zhuhai Key Laboratory of Trusted Large Language Models}
  \city{Zhuhai}
  \country{China}
}
\email{zhzibin@mail.sysu.edu.cn}

\authorsaddresses{
  Mingwei Liu, Zheng Pei, Yanlin Wang, Zihao Wang, Zikang Li, Enci Lin, and Zibin Zheng are with the School of Software Engineering and the Zhuhai Key Laboratory of Trusted Large Language Models, Sun Yat-sen University, Zhuhai, China.
  Email: liumw26@mail.sysu.edu.cn, peizh3@mail2.sysu.edu.cn, wangylin36@mail.sysu.edu.cn, wangzh778@mail2.sysu.edu.cn, lizk26@mail2.sysu.edu.cn, linenc@mail2.sysu.edu.cn, zhzibin@mail.sysu.edu.cn. \\
  Xin Peng is with Fudan University, Shanghai, China.
  Email: pengxin@fudan.edu.cn.
}

\begin{color}{red}
\begin{abstract}
In low-resource framework development (e.g., HarmonyOS), large language models (LLMs) often lack sufficient pre-training exposure, resulting in poor code generation performance. Although they generally preserve programming logic across languages, they frequently fail on framework-specific APIs and syntax, revealing a gap between learned algorithmic knowledge and unfamiliar framework conventions. Consequently, even advanced models such as GPT-4o struggle to produce correct code without prior exposure.

Inspired by these challenges, we propose \app, a framework that leverages API knowledge graphs to synthesize API-oriented question–code pairs without requiring executable environments. It incorporates both single-API and multi-API information, with the latter guided by uncertainty estimation (UE) and Monte Carlo Tree Search (MCTS), to construct high-quality fine-tuning data. For evaluation, we select HarmonyOS as a case study due to its accessible documentation and growing ecosystem, and build the first benchmark for its code generation. Experimental results show that fine-tuning Qwen2.5-Coder-7B with \app achieves a pass@1 of 25.00\%, outperforming untuned GPT-4o (17.59\%). We further observe that larger volumes of data generated by \app consistently lead to better fine-tuning performance, and that the optimal Single-API to Multi-API ratio is 8:2. Ablation studies also confirm the necessity and effectiveness of each component in our framework. These findings highlight the effectiveness of API-oriented data in enhancing LLM performance for low-resource software development scenarios.
\end{abstract}

\end{color}



\maketitle

\section{Introduction}\label{sec:intro}
\begin{figure}[htbp]
    \centering
    \includegraphics[width=0.6\columnwidth]{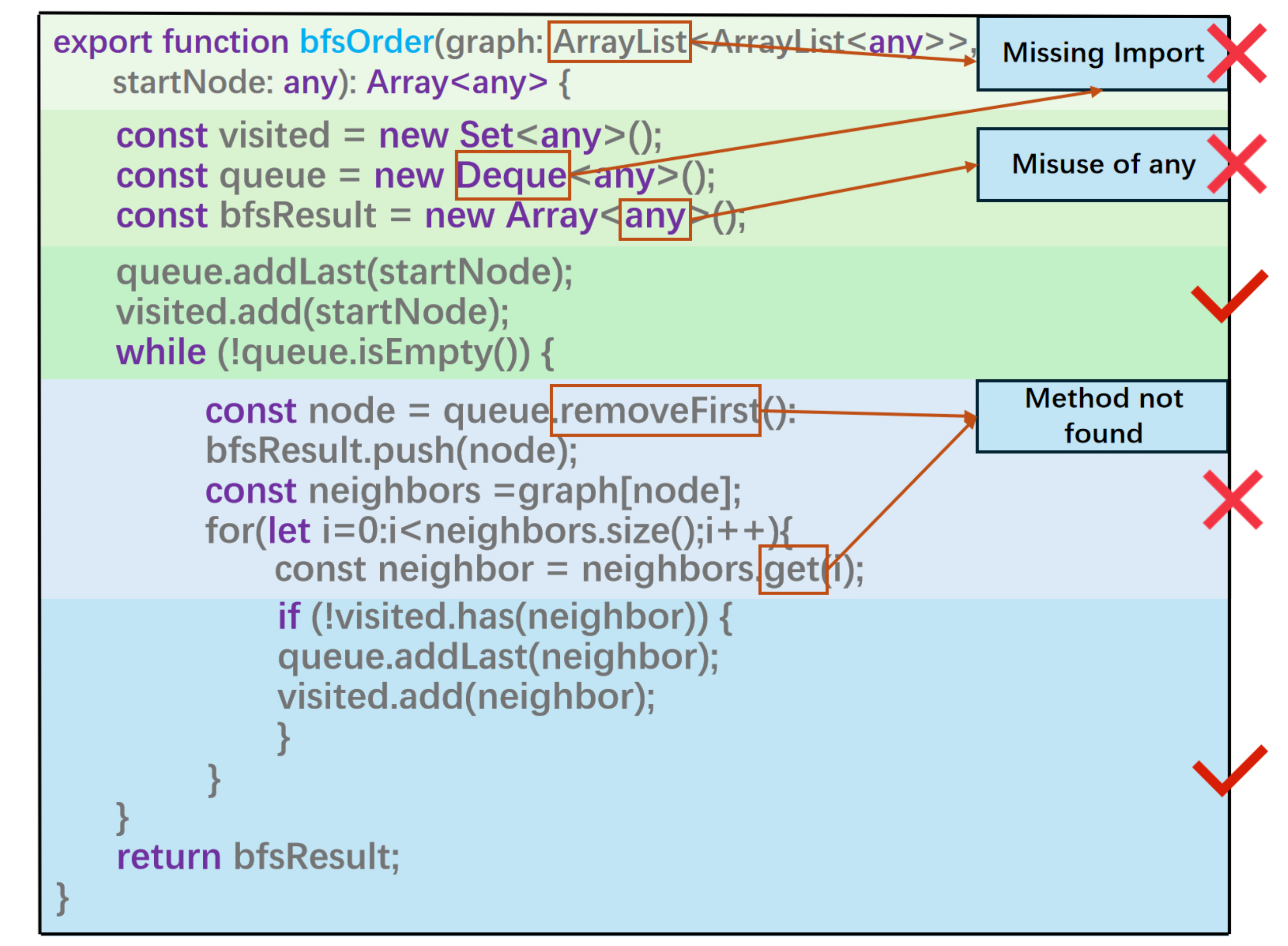}
    \caption{The breadth-first traversal algorithm code under HarmonyOS generated by GPT-4o}
    \label{fig:example_wrong}
\end{figure}

LLMs have achieved remarkable success in software development tasks such as code completion, generation, and comprehension ~\cite{LiJia2023SKCODER,di2025enhancingcodegenerationbidirectional,tian2024fixinglargelanguagemodels,jiang2025rocodeintegratingbacktrackingmechanism,lin2024soen101codegenerationemulating}. However, in emerging frameworks or domain-specific environments, their performance often degrades due to limited exposure during pretraining~\cite{wang2025exploracoderadvancingcodegeneration}. Nonetheless, low-resource scenarios—including private libraries, proprietary frameworks, and other settings with limited data—are commonly encountered in industrial practice. HarmonyOS, a relatively new operating system with distinctive APIs, exemplifies such a low-resource scenario. Developers working with HarmonyOS face unique challenges due to complex API structures and sparse training data, which restrict the model’s ability to generate accurate and contextually appropriate code~\cite{zan-etal-2022-language}. As illustrated in Figure~\ref{fig:example_wrong}, this example shows the code generated by the GPT-4o model for a breadth-first traversal task in the HarmonyOS. Although the generated code is logically correct, it omits the required API import statements, employs the disallowed any type in HarmonyOS, and calls methods that do not exist in ArrayList. \textbf{This observation indicates that the model has internalized general algorithmic knowledge during pretraining, yet it lacks domain-specific knowledge for low-resource scenarios such as HarmonyOS.}

For low-resource programming languages, LLMs often lack familiarity with their syntax and APIs. In contrast, they typically demonstrate stronger competence in code logic. Pan ~\cite{Pan_2024} et al. observed that in code translation tasks, when two languages are used to solve the same problem, models tend to preserve logical components such as control flow and algorithmic structures, while frequently introducing errors in API mapping and syntactic details. This is because, across languages such as C, Python, and Java, the solutions to a given problem are generally consistent in logic, and language-specific features rarely invalidate the overall problem-solving approach. Similarly, Alessandro Giagnorio ~\cite{giagnorio2025enhancingcodegenerationlowresource} et al. reported that models perform poorly in API mapping across programming frameworks, further indicating that LLMs exhibit uneven familiarity with APIs across different languages, although their understanding of code logic remains relatively clear. 

Although domain-specific fine-tuning improves model performance, obtaining high-quality datasets for new platforms is still difficult. Existing methods, like OSS-Instruct~\cite{wei2024magicoder}, generate training data for supervised fine-tuning (SFT), but often lack complete API semantic coverage, sufficient multi-API reasoning, or rely heavily on real usage examples that are not available for new frameworks. These limitations create gaps in data quality and applicability for low-resource scenarios.

To address these challenges, \textbf{we propose \app, a knowledge-graph-based data synthesis framework specifically designed for domain-specific software development}. Our approach constructs a structured API knowledge graph from documentation, capturing function descriptions, parameters, return values, and hierarchical relationships. LLMs are then leveraged to quantify the familiarity with different API nodes, which in turn guides Monte Carlo Tree Search~\cite{metropolis1949monte} (MCTS) to identify unfamiliar or highly correlated API nodes for generating multi-API joint scenarios. For single-APIs, we also provide representative single-API application scenarios. Based on this information, API-Oriented question–code tuples are synthesized to enrich LLMs’ understanding of APIs in low-resource environments, and even to enable the construction of corresponding models in zero-resource scenarios. In contrast to approaches that rely on high-quality exemplar code~\cite{wei2024magicoder}, \app leverages LLMs with API documentation to generate realistic code without referring to real-world cases, even in unfamiliar ecosystems. This enables scalable data construction while explicitly promoting API reuse to avoid redundancy. The resulting data effectively supports fine-tuning for low-resource languages.

In addition, we construct a HarmonyOS code generation benchmark to quantitatively evaluate LLM performance in this low-resource scenario, providing a standard testbed for future research. We conduct extensive experiments (details in Section 3) to evaluate the effectiveness of our method.  The results demonstrate that fine-tuning LLMs with our synthesized data yields substantial improvements over baseline methods on both single-API and multi-API tasks, without relying on any existing real-world cases. The data ratio and data volume studies further verify the correctness of the performance trends of models fine-tuned with data generated by \app, as well as the effectiveness of our selected data ratio. Moreover, ablation studies confirm the necessity of each component in our framework, and the fine-tuned models consistently outperform the plain text + RAG approach for code generation. Specifically, our approach attains a pass@1 of 25.00\% on HarmonyOS benchmark, exceeding the strongest baseline (10.19\%) by 14.81\%. These findings validate the efficiency and effectiveness of API knowledge graph–guided data synthesis.

In summary, this paper makes the following contributions:
\begin{itemize}
    \item We propose a knowledge-graph-driven data synthesis framework that generates targeted training data without relying on execution examples.
    \item We construct a HarmonyOS code generation benchmark to provide a standard evaluation testbed for future research.
    \item We validate the effectiveness of our method through quantitative experiments, showing improvements in LLM performance for both single-API and multi-API scenarios.
    \item We also release the OHBen Dataset, our generated training dataset, as an open-source contribution to support fine-tuning of HarmonyOS–related LLMs.
\end{itemize}

\section{Method}
\label{sec:method}

LLMs often lack sufficient exposure to APIs in emerging frameworks, 
leading to poor generalization in low-resource settings. 
To address this gap, we design \textbf{\app}, a knowledge-graph-driven data synthesis 
framework that creates realistic, API-grounded training samples for HarmonyOS. 
The approach systematically constructs an API knowledge graph, 
generates single-API questions with factual grounding, 
and extends to multi-API scenarios using Monte Carlo Tree Search (MCTS)~\cite{metropolis1949monte}. 

\begin{figure}[htbp]
    \centering
    \includegraphics[width=1\columnwidth]{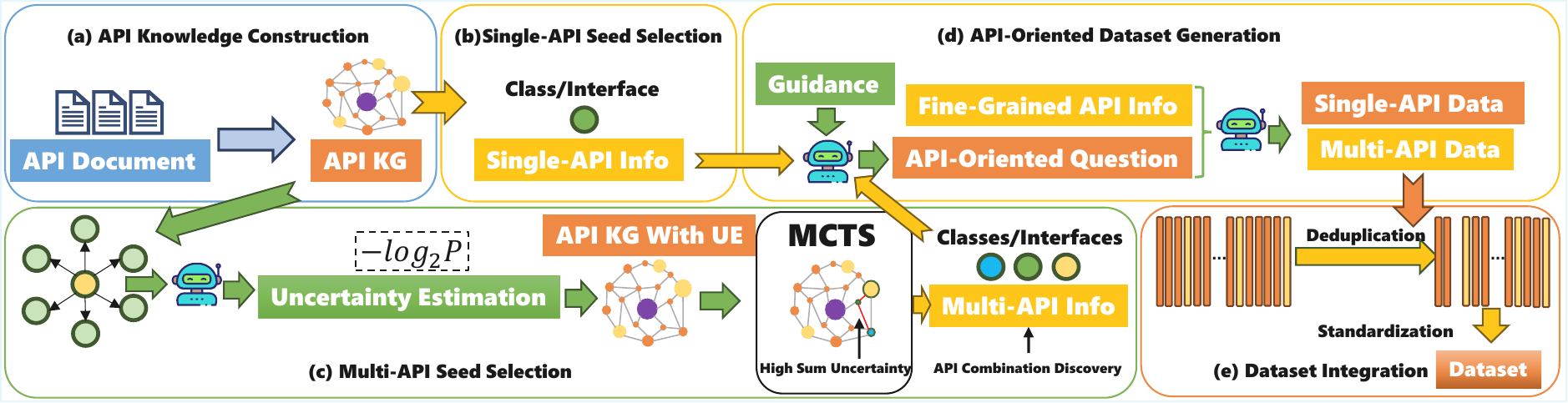}
    \caption{Approach Overview of \app}
    \label{fig:overview}
\end{figure}

Figure~\ref{fig:overview} illustrates the overall workflow. 
The framework operates through five key stages:

\begin{enumerate}
  \item \textbf{API Knowledge Graph Construction:} 
  HarmonyOS API documentation is parsed to extract modules, classes, methods, and properties, 
  along with hierarchical and semantic relationships. 
  Each node is enriched with metadata such as functional description, parameters, and return values.

    \item \textbf{Single-API Seed Selection:} 
    By leveraging the API knowledge graph, the system retrieves and provides the corresponding single-API information based on a specific usage scenario.
    \item \textbf{Multi-API Seed Selection:} 
    Leveraging the API knowledge graph, we employ LLMs to compute UE, which guides a Monte Carlo Tree Search (MCTS~\cite{metropolis1949monte}) to identify node paths with the highest cumulative uncertainty. The relevant nodes along these paths are then extracted and combined to construct multi-API information.

    \item \textbf{API-Oriented Question-Code Tuples Dataset Construction:} 
    Using the provided single-API and multi-API information, LLMs generate appropriately tailored API-Oriented problem scenarios. The models then retrieve fine-grained API information from the API knowledge graph to produce corresponding API-Oriented code solutions, thereby constructing an API-Oriented question–code tuple dataset.

     \item \textbf{Dataset Integration and Post-processing:} 
    The single-API and multi-API generated API-Oriented data are combined into a unified dataset. 
    Post-processing steps include deduplication, formatting, and validation of questions and answers, 
    ensuring consistency and quality for downstream fine-tuning tasks.
\end{enumerate}

These five steps form the core of the \app methodology. 
Details regarding implementation, dataset construction, and descriptive statistics 
are provided in Section~\ref{sec:implementation}.

\subsection{API Knowledge Graph Construction}
In this stage, we parse the API documentation of low-resource frameworks (HarmonyOS) to build a structured knowledge graph representing the API ecosystem. The graph captures hierarchical relationships among modules, classes, methods, and properties. Each node includes metadata such as functional descriptions, parameters, and return values. This structured representation forms the basis for generating both single-API and multi-API training data. The construction process involves:
\begin{itemize}
    \item Extract code information and text information from the API documentation.
    \item Transform restructured code and semantic information into a professional API knowledge graph.
\end{itemize}
\begin{figure}[htbp]
  \centering
  \includegraphics[width=1\columnwidth]{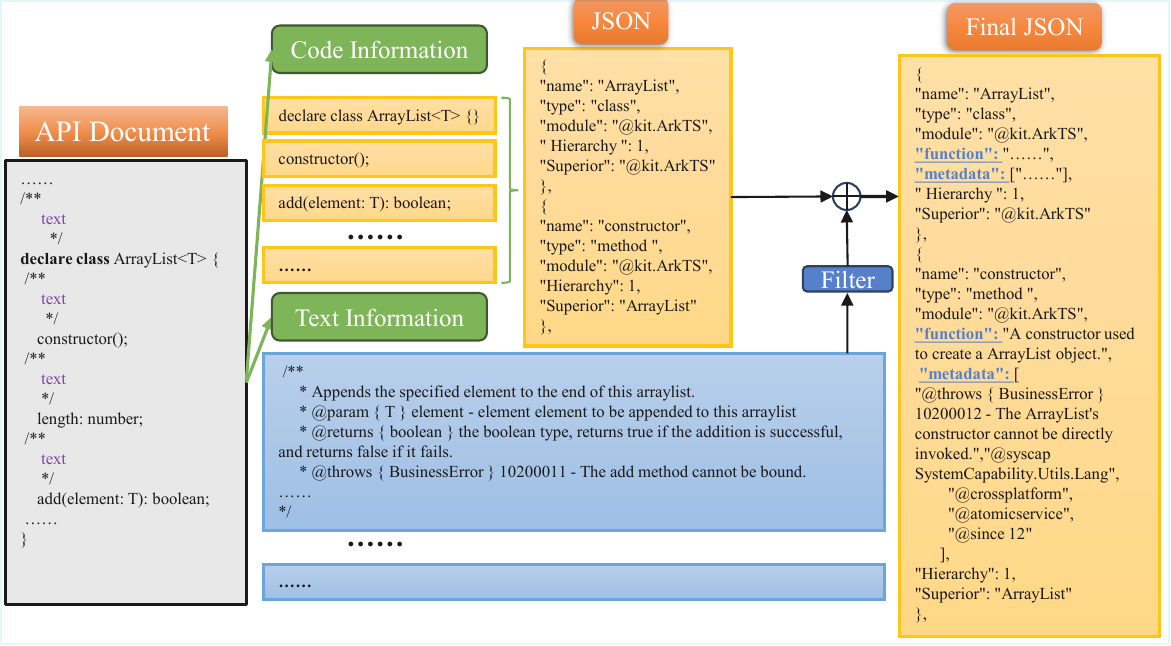}
  \caption{API Documentation Conversion Workflow}
  \label{fig:file}
\end{figure}

\parabf{Data Reorganization.}
As illustrated in Figure~\ref{fig:file}, we extracted all API documentation from the SDK folders of DevEco Studio, the official IDE for HarmonyOS. We scanned the entire set of API documents and applied rule-based matching to extract information from each line. The extracted data typically comprise two categories: code-related information and textual information.

\parabf{Code Info.}
The code-related information contains all code snippets from the API documentation, which inherently encode the graph structure of the APIs. By analyzing these code snippets, we can determine the hierarchical organization of different methods, properties, classes, interfaces, namespaces, and enums.

\parabf{Text Info.}
The textual information includes metadata for the corresponding APIs, such as functional descriptions, parameters, and return values. This information can be leveraged by large language models as reliable evidence for subsequent code generation.

\begin{figure}[htbp]
  \centering
  \includegraphics[width=0.8\columnwidth]{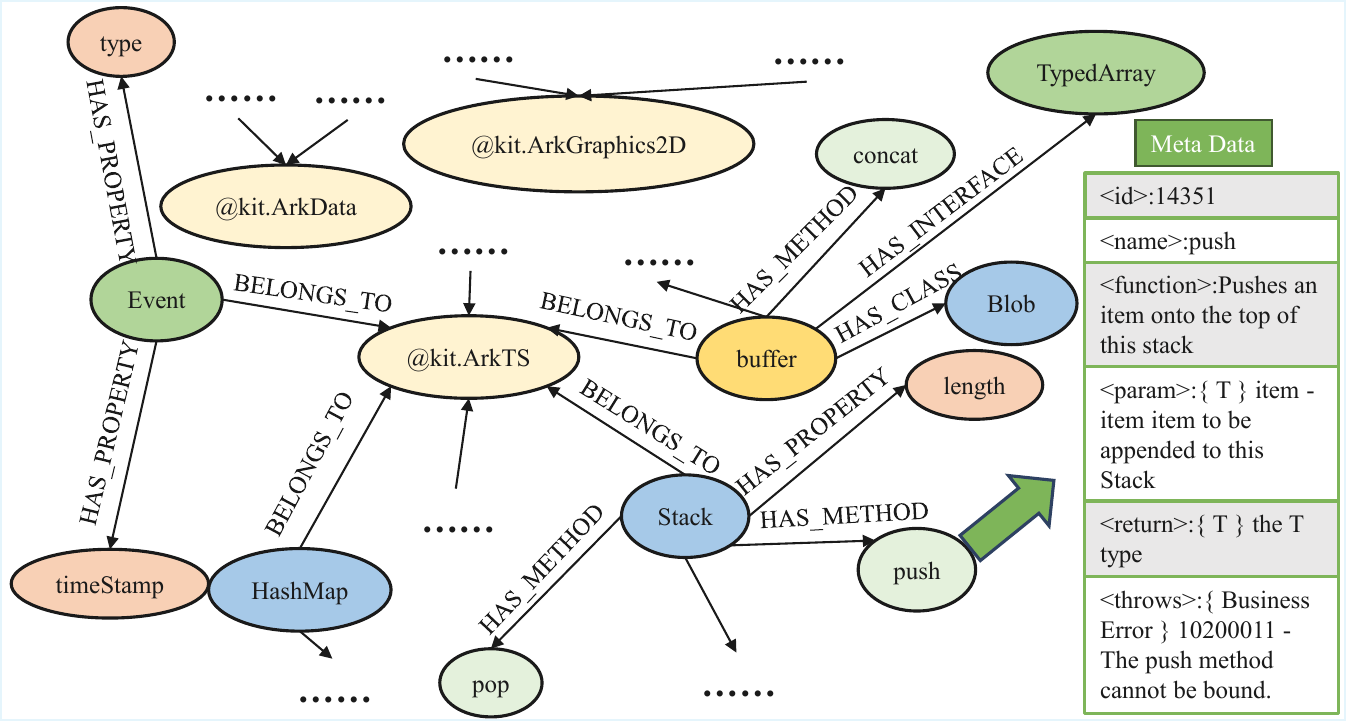}
  \caption{API Knowledge Graph and Node Metadata}
  \label{fig:APIKG}
\end{figure}

\parabf{API Knowledge Graph Construction.}
All entities are extracted from the code and their relationships are formed based on nesting structures to create an initial graph. The corresponding textual information is parsed, with rule-based matching and filtering to get the most up-to-date and valid semantic data. By combining this semantic data with the initial graph structure, as shown in Figure~\ref{fig:APIKG}, an API knowledge graph is created.

\subsection{Single-API Seed Selection}
\begin{figure}[htbp]
    \centering
    \includegraphics[width=0.5\textwidth]{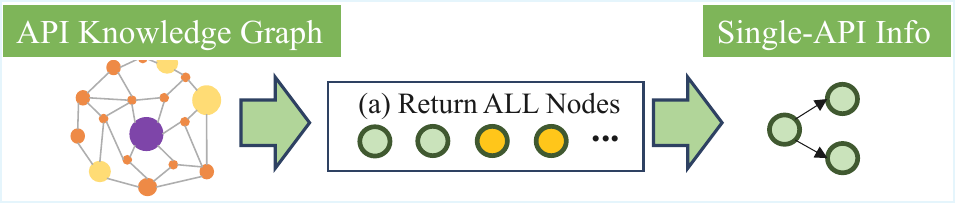}
    \caption{Single-API Non-Leaf Node Selection and Data Retrieval}
    \label{fig:single_api}
\end{figure}

At this stage, the \app framework traverses all non-leaf nodes in the knowledge graph, such as classes and interfaces, and retrieves the semantic and code information of these nodes along with their child nodes for output. The selection of single-API seeds is illustrated in Figure~\ref{fig:single_api}.

\subsection{Multi-API Seed Selection}
At this stage, for multi-API information, we first employ a large language model to compute the UE for each non-leaf node, thereby quantifying the model’s unfamiliarity with the node. This UE then guides a Monte Carlo Tree Search (MCTS)~\cite{metropolis1949monte} to identify the top-5 highest-scoring paths. From each path, which contains several high-scoring nodes, 2–3 nodes are randomly sampled and combined to construct multi-API information.
The detailed procedure for synthesizing multi-API information is as follows:
\begin{itemize}
    \item Compute the UE for each non-leaf node.
    \item MCTS~\cite{metropolis1949monte} is executed with the UE value serving as the reward function.
\end{itemize}

\parabf{Uncertainty Estimation Construction.}
Shannon (1948) proposed that uncertainty estimation~\cite{Shannon} quantifies the information content of a "fact" under a given probability distribution. In our API knowledge graph, a "fact" is represented by a non-leaf node, its code and semantic information, and its child nodes. This fact is formalized as a triplet $\tau = <u, \rho, v>$. To measure how well a large language model understands this information, we transform the triplets into a form for computing conditional probabilities, specifically $P(v \mid u, \rho)$. Given a head entity $u$ and a predicate $\rho$, we estimate the probability that a tail entity $v$ belongs to $u$. The negative logarithm of this probability gives the uncertainty estimation, calculated using the following formula:

\begin{equation}I(u,\rho,v) = -\log_2 P(v \mid u,\rho)\end{equation}

Let $u$ denote a non-leaf node, such as the $ArrayList$ class, and $v$ denote one of its child nodes, such as the $add$ method. The relation $\rho$ represents the membership of $v$ under $u$, and in our knowledge graph, $\rho$ is modeled as a directed relationship.$P(v \mid u,\rho)$ thus represents the probability of 
$v$ belonging to $u$ given the leaf node, while $I(u,\rho,v)$ denotes the computed uncertainty estimation. This value reflects the extent to which a large language model understands the internal structure of the upper-level API, providing a quantifiable measure that can be used in further analyses. Finally, the computed uncertainty estimation value is stored as a property of the corresponding node.

\begin{figure}[htbp]
    \centering
    \includegraphics[width=0.7\textwidth]{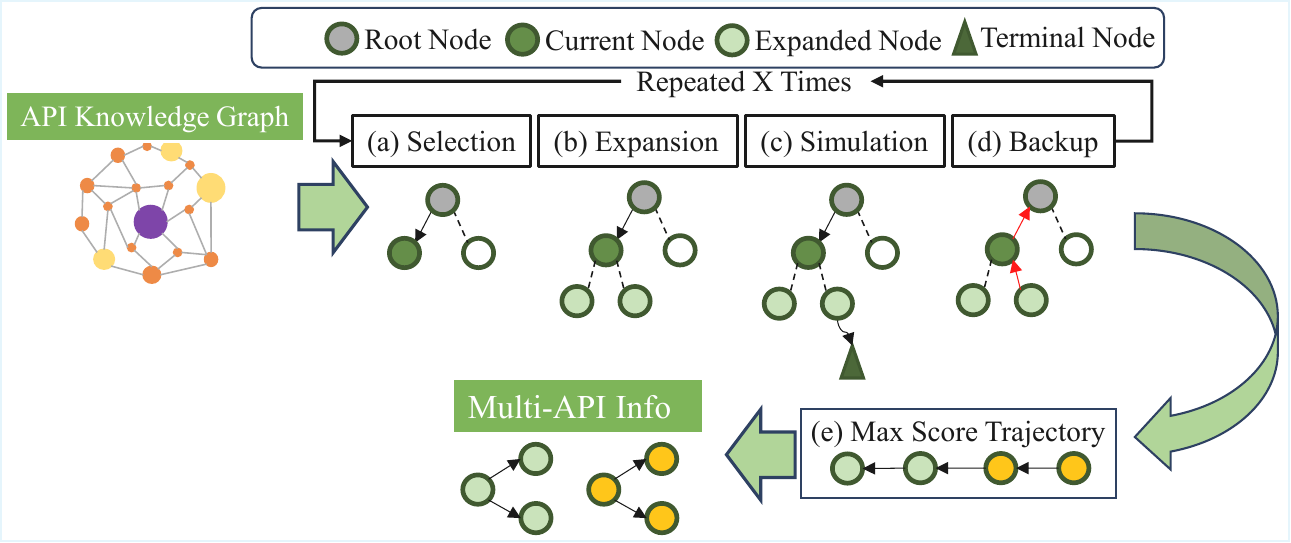}
    \caption{Multi-API Non-Leaf Node Selection and Data Retrieval}
    \label{fig:multi_api}
\end{figure}

\parabf{MCTS.}
The overall workflow of integrating the MCTS~\cite{metropolis1949monte} algorithm into our \app framework is illustrated in Figure~\ref{fig:multi_api}.

\textbf{Node Selection.}
Starting from a root node, the algorithm recursively selects child nodes according to the chosen selection strategy until a leaf node or an unexpanded node is reached. In our implementation, we adopt the Upper Confidence Bound (UCB1~\cite{Auer2002}) strategy. Importantly, the root node is not fixed; instead, each node is treated as a root once, aiming to explore a broader range of API combination paths. The UCB1~\cite{Auer2002} strategy is defined as follows:
\begin{equation}
\text{UCB1}(n) =
\begin{cases}
+\infty, & N(n) = 0, \\[4pt]
Q(n) + c \cdot \sqrt{\dfrac{\ln N_{\text{total}}}{N(n)}}, & N(n) > 0
\end{cases}
\end{equation}

\begin{equation}
Q(n) = \frac{W(n)}{N(n)}
\end{equation}

In the formula, $n$ is the current node, $N(n)$ is the visit count of node $n$, and $W(n)$ is its total accumulated reward. In our implementation, the reward for each node is the uncertainty estimation from Section 2.1. $Q(n) = W(n)/N(n)$ represents the average reward of the node. The parameter $c$ is the exploration coefficient, balancing exploration and exploitation. $N_{\text{total}}$ is the total visit count of the parent node or the sum of all sibling nodes' visits. The value $+\infty$ ensures unvisited nodes are selected at least once.

\textbf{Expansion.}
In the expansion step, if the selected leaf node has unvisited neighbors, one is randomly chosen and added as a new child node, expanding the tree and exploring new states.

\textbf{Simulation.}
After expansion, a simulation starts from the newly added node. The algorithm randomly selects unvisited successor nodes until a terminal node or a leaf node with no unvisited successors is reached. The simulated path is denoted as:
\begin{equation}
\pi = (n_0, n_1, \dots, n_T),
\end{equation}
where $n_0$ is the newly expanded node and $n_T$ is the terminal node. The cumulative reward of this path is calculated as:

\begin{equation}
R(\pi) = \sum_{t=0}^{T} r(n_t),
\end{equation}

where $r(n_t)$ represents the reward of node $n_t$. In our implementation, the reward for each node is defined as the uncertainty estimation introduced in Section 2.1. This simulation step provides an estimate of the potential value of the newly expanded node, guiding subsequent backpropagation and selection.

\textbf{Backup.}
After simulation, the reward obtained from the simulated path is propagated back along the nodes traversed during selection and expansion. For each node $n$ in the path $\pi = (n_0, n_1, \dots, n_T)$, the visit count and total accumulated reward are updated as follows:
\begin{equation}
N(n) \leftarrow N(n) + 1, \quad W(n) \leftarrow W(n) + R(\pi), \quad Q(n) = \frac{W(n)}{N(n)},
\end{equation}
where $R(\pi)$ is the cumulative reward of the simulated path. This backpropagation step refines the estimated value of each node, influencing future selection decisions and guiding the search towards nodes with higher expected rewards.

\textbf{Max Score Trajectory.}
After repeating the process multiple times, we obtain several paths. We select the top five highest-scoring paths, each containing multiple non-leaf API nodes. From these nodes, we randomly select two to three API nodes, and return their corresponding code and semantic information. This procedure provides multi-API nodes along with their code and semantic data.

\subsection{API-Oriented Question-Code Tuples Generation}
This stage focuses on the generation of an API-Oriented question–code tuple dataset. Specifically, we leverage the code information and semantic descriptions of API nodes retrieved from the API knowledge graph. By utilizing these reliable sources in combination with carefully designed prompts, the system generates well-structured, low-resource framework-related questions. Subsequently, finer-grained API information is retrieved based on these questions and provided to LLMs, which then produce the corresponding code solutions. The key steps include:
\begin{itemize}
    \item \textbf{Target API Information Retrieval:} Select non-leaf nodes as targets and provide reliable code information and semantic information.  
    \item \textbf{API-Oriented Questions Generation:} LLMs are employed to generate well-structured low-resource framework-related questions grounded in the provided API information.  
    \item \textbf{API-Oriented Code Generation:} LLMs address low-resource framework-related questions by leveraging the questions themselves together with fine-grained API information retrieved from the API knowledge graph.
\end{itemize}

\begin{figure}[htbp]
  \centering
  \includegraphics[width=1\columnwidth]{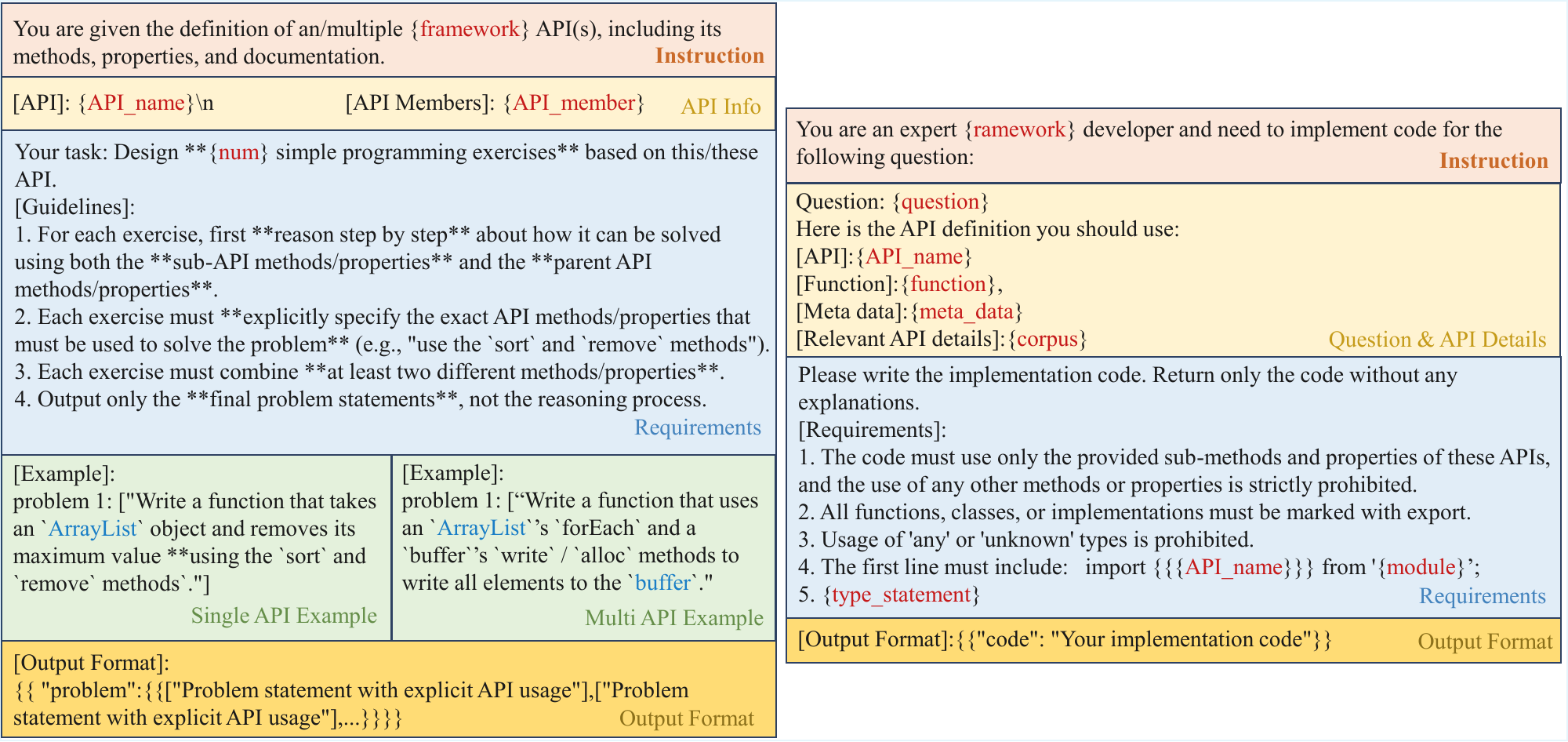}
  \caption{Left: Prompt for question generation. Right: Prompt for code generation.}
  \label{fig:prompt_all_gen}
\end{figure}

\parabf{Prompting For Generating Question.}
For both single-API and multi-API data generation, the previous step provides the code and semantic information of the target nodes, as well as the code and semantic information of their corresponding child APIs. With this information, we populate a specially designed prompt for question generation. This prompt is tailored to generate programming questions in the low-resource framework environment and is capable of producing questions involving single-API as well as multi-API usage, as illustrated in Figure~\ref{fig:prompt_all_gen}

\parabf{Prompting For Generating Code.}
Once low-resource framework-related questions are formulated, LLMs are employed to solve them. To enhance problem-solving, the \app approach retrieves fine-grained code and semantic information from the knowledge graph for the relevant single-API or multi-API elements mentioned in each question. This retrieved information is incorporated into a carefully designed code-generation prompt, as illustrated in Figure~\ref{fig:prompt_all_gen}. By leveraging this enriched knowledge, the model gains a comprehensive understanding of how each API is used and the functionalities they provide, thereby enabling more effective solutions to low-resource framework-related problems.
\subsection{Dataset Integration and Post-processing}
\begin{figure}[htbp]
  \centering
  \includegraphics[width=0.5\columnwidth]{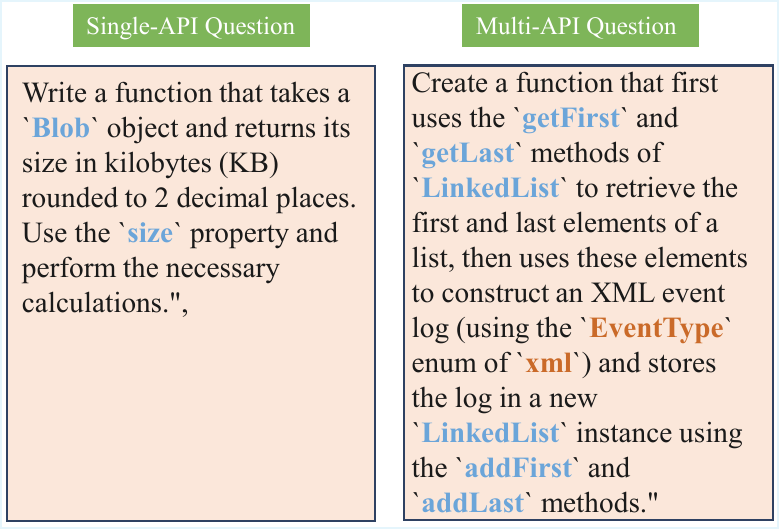}
  \caption{Single-API Scenario Questions and Multi-API Scenario Questions}
  \label{fig:data_example}
\end{figure}

Dataset integration and post-processing occur after the generation of both single-API and multi-API data. In total, our method produces 6,400 single-API data points and 1,600 multi-API data points. The generated API-oriented questions of different types are shown in Figure~\ref{fig:data_example}.

The generated datasets are then merged and subjected to post-processing to ensure consistency and quality. This step includes:
\begin{itemize}
    \item Deduplication of questions and answers.
    \item Standardizing the format of dataset entries.
    \item Preparing the dataset for downstream training and evaluation.
\end{itemize}

\parabf{Deduplication.}
To ensure that the generated dataset does not contain samples similar to those in the benchmark (i.e., with similar questions or solutions), we employ a fuzzy text-matching algorithm based on the Levenshtein distance to identify and remove similar samples. Equivalent numbers of new samples are then generated to maintain the dataset size, thereby preserving the rigor and consistency of the experiments.

\parabf{Standardization.}
Our original data format does not conform to the training format required by LLaMA-Factory~\cite{zheng2024llamafactory}. Therefore, we standardize the dataset into the format adopted by LLaMA-Factory~\cite{zheng2024llamafactory}, to facilitate subsequent fine-tuning.

\subsection{Implementations}
\label{sec:implementation}
The \textbf{\app} framework is implemented in Python and leverages the 
\textbf{DeepSeek-V3} LLM for question and answer synthesis. 
Knowledge graph construction, MCTS~\cite{metropolis1949monte}, and data integration are automated to produce 
a structured dataset in JSON format. Each record contains:
\begin{itemize}
    \item \texttt{question}: a natural-language programming query.
    \item \texttt{code}: corresponding explanation or code snippet.
    \item \texttt{api\_nodes}: the set of API nodes involved in generating the sample.
\end{itemize}
\parabf{Model Selection.}

\textbf{Deepseek-V3.}
Deepseek-V3~\cite{deepseekai2025deepseekv3technicalreport} is selected to generate uncertainty estimation values, low-resource framework-related questions, and their corresponding code solutions. The choice of Deepseek-V3~\cite{deepseekai2025deepseekv3technicalreport} is motivated by its strong coding capabilities, superior reasoning ability, and relatively low cost, making it a suitable model within our methodological pipeline.

\textbf{Fine-tuned LLMs}.
We select \qwentwo-7B, \mistral-7B, and \deepseek-7B as the LLMs for fine-tuning, as they are relatively lightweight in terms of parameter size while still exhibiting strong coding capabilities. In addition, \mistral-7B is a model fine-tuned with the OSS-Instruct~\cite{wei2024magicoder} approach, which further motivated our interest in including it in our study.

\textbf{LLMs used for comparison.}
We select \qwenthree-32B, \gpt, \omini, and \deepseek-671B as LLMs for comparison. \qwenthree-32B is the upgraded version of \qwentwo-7B, and we specifically choose the 32B variant to facilitate a direct comparison between the fine-tuned \qwentwo and its successor. \deepseek-671B represents the full-scale version of \deepseek-7B, which further motivates our interest. In addition, we include \gpt and \omini from OpenAI: \gpt is known for its exceptional capabilities and stands among the leading commercial closed-source models, while \omini, a smaller-scale variant released by OpenAI, is chosen to assess the performance of lightweight models from the same provider.

\parabf{Fine-tune configuration.}
For hardware, we used an RTX 4090 GPU for fine-tuning and an Intel Xeon Platinum 8352V CPU running Linux. The fine-tuning platform was LLaMA-Factory~\cite{zheng2024llamafactory}, a comprehensive framework for LLM fine-tuning. All models in our experiments were fine-tuned using this platform. We applied the Lora method, training for 6 epochs with the AdamW optimizer, an initial learning rate of 5e-5, and a batch size of 2.

\section{Evaluation}\label{sec:evaluation}

We conduct extensive experiments to evaluate the effectiveness of \app. Specifically, we aim to answer the following research questions:

\begin{itemize}
  \item \textbf{RQ1 (Effectiveness):} Is the \app method more effective than data synthesized using the OSS-Instruct~\cite{wei2024magicoder} method?
  \item \textbf{RQ2 (Data Scale Impact):} The scale of synthetic data and the ratio between independent API code data and linked API code data influence the fine-tuning of LLMs.
\item \textbf{RQ3 (Single-API Data Vs. Multi-API Data):} How does single-API Data compare with multi-API Data in enhancing the fine-tuning effectiveness of LLMs?
\item \textbf{RQ4 (Component Effectiveness):} How effective are the individual components of the \app method?
\end{itemize}

\subsection{Experimental Setup}

\begin{figure}[htbp]
  \centering
  \includegraphics[width=0.8\columnwidth]{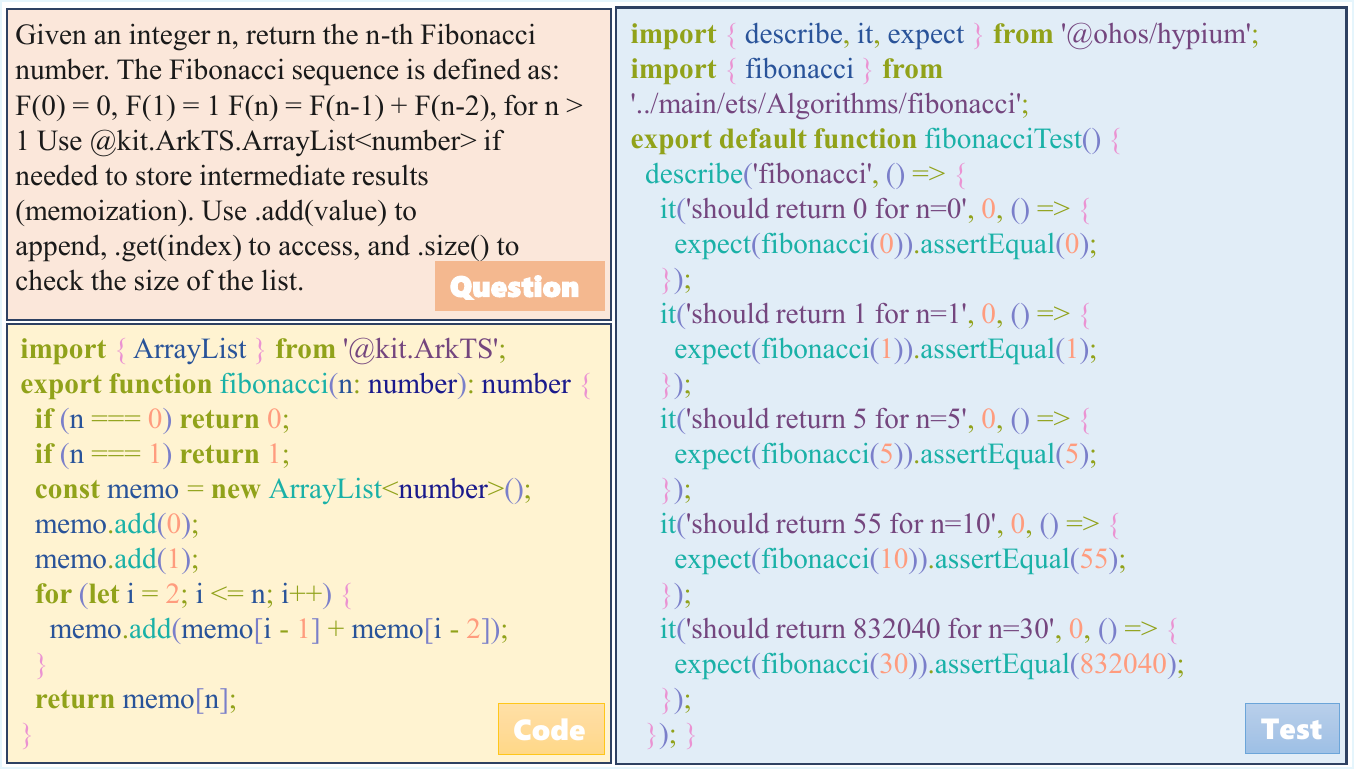}
  \caption{An example from the OHBen}
  \label{fig:benchmark_example}
\end{figure}

\begin{figure}[htbp]
  \centering
  \includegraphics[width=0.6\columnwidth]{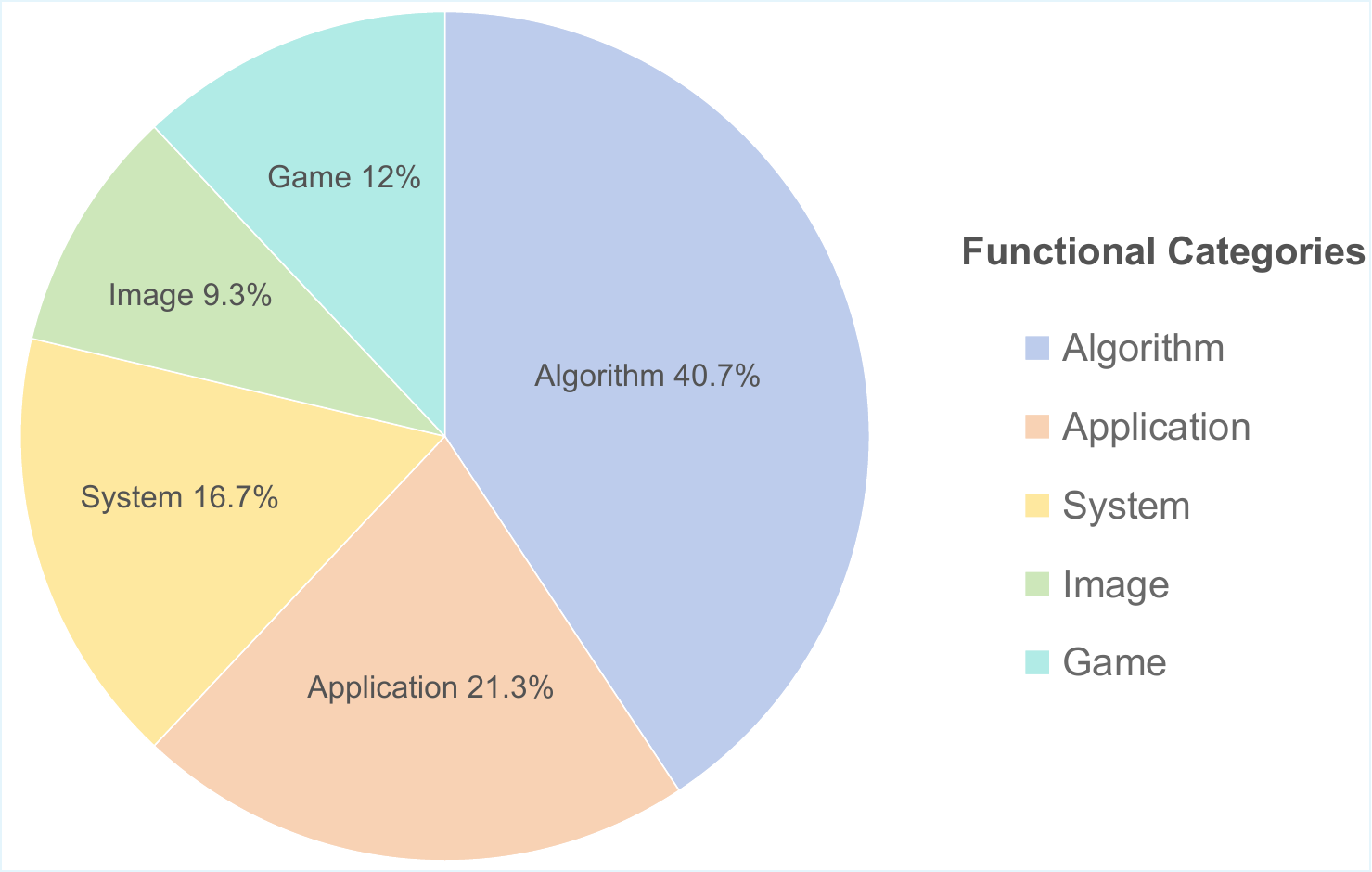}
  \caption{Distribution of HarmonyOS benchmark sample categories}
  \label{fig:dataset}
\end{figure}

\parabf{Datasets.} In this study, we introduce the first dedicated HarmonyOS benchmark designed to assess the proficiency of LLMs in HarmonyOS programming. Considering that existing low-resource frameworks such as Qt, ROS, and HarmonyOS lack publicly available benchmark, and given the relatively high accessibility of HarmonyOS documentation, we select HarmonyOS as the target framework for evaluation. As shown in the Figure~\ref{fig:dataset}, we devoted approximately 300 human-hours to manually curate a comprehensive suite of tasks covering algorithms, graphics, gaming, and audio. Each task includes minimal yet essential test cases to rigorously evaluate code correctness and logical consistency. Specifically, for each sample in the benchmark, as illustrated in Figure~\ref{fig:benchmark_example}, it comprises a problem that requires a solution using HarmonyOS APIs, a corresponding code implementation, and a complete test script.

\begin{figure}[htbp]
  \centering
  \includegraphics[width=0.5\columnwidth]{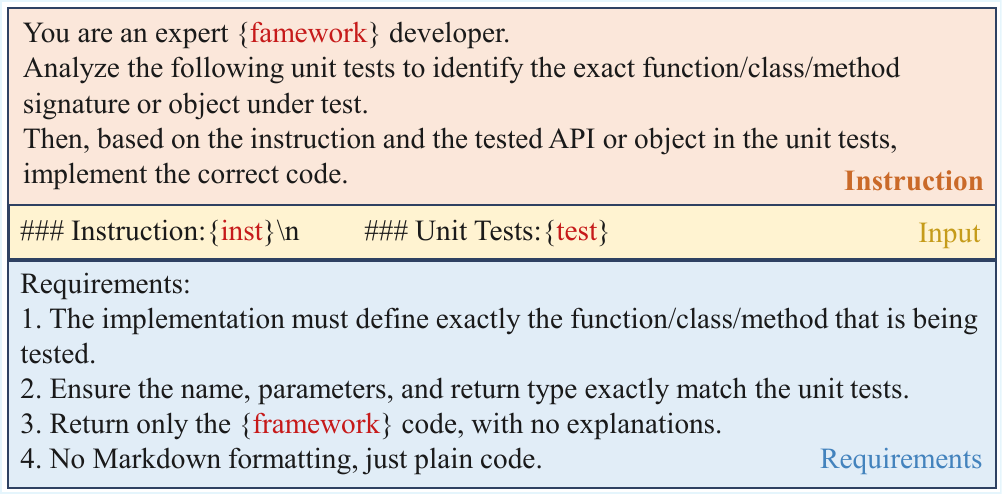}
  \caption{Prompt Used for Evaluating HarmonyOS Benchmark}
  \label{fig:eval_prompt}
\end{figure}

Moreover, when executing the benchmark, we employ the prompt illustrated in Figure ~\ref{fig:eval_prompt}, which facilitates the construction of testable code based on the unit tests to verify whether the implementation passes.

As shown in Figure~\ref{fig:dataset}, when constructing the HarmonyOS benchmark, 40.7\% of the samples belong to the algorithm category, which is higher than any other category. This predominance is due to the fact that algorithm-related tasks involve numerous fundamental APIs, such as \textbf{Stack}, \textbf{HashMap}, and \textbf{ArrayList}, which are essential for development. Therefore, we intentionally included a larger proportion of algorithmic tasks to evaluate models’ understanding of core APIs.

To better highlight the models’ capability in applying the HarmonyOS for practical application development, we also designed a substantial number of application-oriented tasks, accounting for approximately 17.6\% of the benchmark. Subsequent categories include system, game, and graphics-related tasks, which reflect common scenarios in HarmonyOS development.

\parabf{Baselines.} We selected OSS-Instruct~\cite{wei2024magicoder} as a baseline to compare with our proposed \app method. OSS-Instruct~\cite{wei2024magicoder} is a highly effective and well-established program synthesis approach, renowned for its ability to fine-tune 7-billion-parameter models using synthetic data with strong performance. This method represents a mature and robust reference in code generation and is widely adopted in the field of code data synthesis. To the best of our knowledge, there are currently no other methods specifically leveraging code synthesis data for model fine-tuning. Therefore, we select OSS-Instruct~\cite{wei2024magicoder} as the sole baseline for comparison.

When using OSS-Instruct~\cite{wei2024magicoder}, the method relies on a so-called heuristic dataset. However, to date, no publicly available HarmonyOS-related datasets are known. To ensure a fair comparison, we crawled 19,070 code snippets from the HarmonyOS documentation to construct a heuristic dataset, thereby adapting OSS-Instruct~\cite{wei2024magicoder} for HarmonyOS-specific tasks.

\parabf{LLM.}
For fine-tuning, we selected the 7B-parameter versions of \deepseek-7B, \qwentwo-7B, and \mistral-7B. The 7B scale was chosen as it represents a relatively efficient configuration that balances computational feasibility with experimental rigor. Both \deepseek-7B and \qwentwo-7B have gained recognition for their strong code generation capabilities, making them natural candidates for our experiments. In addition, \mistral-7B was included since it serves as a representative model trained with the OSS-Instruct~\cite{wei2024magicoder} method, which we adopt as our baseline; therefore, it was selected as one of the fine-tuned models.

For LLMs without fine-tuning, we evaluated \gpt, \omini, \deepseek-671B, and \qwenthree-32B. \gpt was chosen as a powerful commercial LLM with strong reasoning abilities, while \omini, a lightweight model from OpenAI, also provides competitive reasoning capacity. To benchmark the fine-tuned \deepseek-7B against its full-scale counterpart, we included \deepseek-671B. Similarly, to assess the fine-tuned \qwentwo-7B in comparison with a larger variant, we evaluated \qwenthree-32B.

\subsection{RQ1: Effectiveness}
\parabf{Design.}
Table~\ref{tab:SFT_model_compare} reports a comparative analysis of fine-tuning results obtained using synthetic data generated by our \app method and the OSS-Instruct approach~\cite{wei2024magicoder}, evaluated across a diverse set of LLMs, including both commercial and open-source variants. The models under consideration comprise Qwen-7B, Qwen2.5-7B, \qwentwo-7B, Deepseek-LLM-7B, \deepseek-7B, \mistral-7B, \deepseek-671B, \qwenthree-32B, \gpt, and \omini. Owing to computational resource constraints, fine-tuning experiments were conducted exclusively on the 7B-parameter versions of Qwen, Qwen2.5, \qwentwo, Deepseek-LLM, \deepseek, and \mistral. Given its consistently strong performance across a wide range of tasks, \gpt was selected as the representative commercial LLM. In addition, \omini was included as a lightweight GPT variant to assess its effectiveness on the HarmonyOS benchmark. Among open-source LLMs, \qwenthree~\cite{qwen3technicalreport} and \deepseek have demonstrated particularly competitive capabilities, and were therefore chosen as representative models in this category.

We evaluated 7B models (\deepseek-7B, \mistral-7B, \qwentwo-7B) zero-shot on the HarmonyOS benchmark. To assess synthetic data quality, we fine-tuned these models using OSS-Instruct~\cite{wei2024magicoder} and \app-generated data. Additionally, to investigate model evolution, we tested DeepSeek-LLM-7B, Qwen-7B, and Qwen2.5-7B under zero-shot and \app fine-tuning settings (excluding OSS-Instruct to isolate iteration effects). Finally, commercial models (\gpt, \omini, \deepseek-671B) served as zero-shot references.

\parabf{Results.}
\begin{table}[t]
\centering
\caption{Comparison of pass@1 performance on the HarmonyOS benchmark. 
Models fine-tuned with enhanced data using \app and OSS-Instruct are compared against their w/o SFT counterparts.}
\label{tab:SFT_model_compare}
\begin{tabular}{lllc}
\toprule
\textbf{Model} & \textbf{Release Date} & \textbf{Method} & \textbf{pass@1} \\
\midrule

\deepseek-671B      & 2025-01-20 & w/o SFT & 3.70\%  \\
\gpt                & 2024-05-13 & w/o SFT & 17.59\% \\
\omini-7B           & 2025-04-16 & w/o SFT & 9.26\%  \\
\midrule
\qwenthree-32B      & 2025-04-29 & w/o SFT & 9.26\%  \\
\midrule

\multirow{2}{*}{Deepseek-LLM-7B}  
& \multirow{2}{*}{2023-11-29} 
& w/o SFT & 0\%      \\
& & \app  & 7.41\%   \\

\multirow{3}{*}{\deepseek-7B} 
& \multirow{3}{*}{2025-01-20} 
& w/o SFT        & 0\%     \\
& & OSS-Instruct & 0.93\%  \\
& & \app         & 9.26\%  \\
\midrule

\multirow{3}{*}{\mistral-7B}  
& \multirow{3}{*}{2023-09-27} 
& w/o SFT        & 4.63\%  \\
& & OSS-Instruct & 4.63\%  \\
& & \app         & 14.81\% \\
\midrule
\multirow{2}{*}{Qwen-7B}  
& \multirow{2}{*}{2023-08-04} 
& w/o SFT & 1.85\%  \\
& & \app  & 5.56\%  \\

\multirow{2}{*}{Qwen2.5-7B}  
& \multirow{2}{*}{2024-09-18} 
& w/o SFT & 6.48\%  \\
& & \app  & 16.67\% \\

\multirow{3}{*}{\qwentwo-7B}  
& \multirow{3}{*}{2024-11-12} 
& w/o SFT        & 8.33\%  \\
& & OSS-Instruct & 10.19\% \\
& & \app         & \textbf{25.00\%} \\

\bottomrule
\end{tabular}
\end{table}

\textbf{Comparison with the OSS-Instruct~\cite{wei2024magicoder} Method.}
Using the OSS-Instruct~\cite{wei2024magicoder} method, we scanned and extracted all HarmonyOS code blocks from the documentation in the Gitee repositories of HarmonyOS, and organized them into a heuristic library as described in the OSS-Instruct~\cite{wei2024magicoder} framework. To evaluate the fine-tuning performance under an equal data scale, we generated 8000 HarmonyOS samples with OSS-Instruct~\cite{wei2024magicoder}. As shown in Table~\ref{tab:SFT_model_compare}, under the same model architecture, fine-tuning with data generated by the \app method yields an average improvement of 11.11\% in pass@1 across three models, compared with fine-tuning on OSS-Instruct~\cite{wei2024magicoder} data. The improvements are substantial, primarily because, given the same data volume, our synthesized data more closely adhere to API usage conventions and logical patterns. In contrast, OSS-Instruct~\cite{wei2024magicoder} relies on high-quality prompt libraries, which are not available in low-resource environments. This, in turn, further demonstrates the superior effectiveness of our approach, and we will further analyze the results across different models in the following discussion.

\begin{figure}[htbp]
    \centering
    \includegraphics[width=0.8\textwidth]{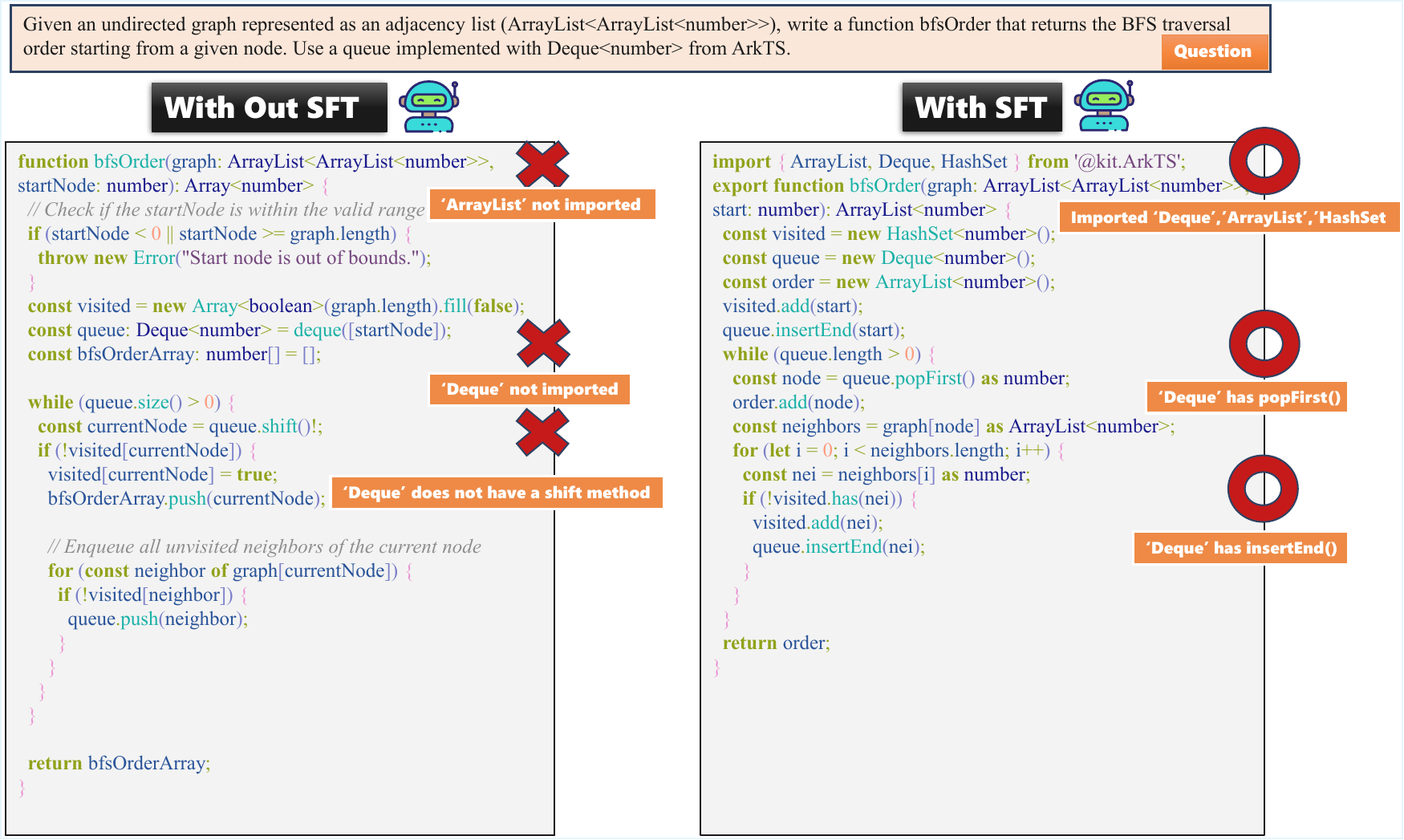}
    \caption{A comparative example illustrating the outputs of the Qwen model on the same problem, with and without fine-tuning}
    \label{fig:rq1_example}
\end{figure}

\textbf{\qwentwo-7B.}
For \qwentwo-7B, the pass@1 score without fine-tuning is 8.33\%, which is the highest among all models prior to fine-tuning. We hypothesize that this may be due to \qwentwo-7B having acquired preliminary exposure to HarmonyOS during pre-training, thereby demonstrating stronger baseline performance compared to other models. When fine-tuned with OSS-Instruct~\cite{wei2024magicoder} data, \qwentwo-7B achieves a modest improvement of 1.86\%. In contrast, fine-tuning with data generated by the \app method results in a substantial improvement of 15.74\%, clearly outperforming OSS-Instruct~\cite{wei2024magicoder}. After fine-tuning, the model demonstrates a deeper understanding of API information. As shown in Figure~\ref{fig:rq1_example}, for the same problem, the fine-tuned model successfully addresses issues such as using APIs without proper import statements and invoking non-existent methods. This provides evidence that fine-tuning enables the model to acquire and apply API-specific knowledge.

\textbf{Qwen-7B \& Qwen2.5-7B.}
For Qwen-7B, pass@1 improves from 1.85\% to 5.56\% after \app-based fine-tuning, indicating a modest gain. For Qwen2.5-7B, it increases from 6.48\% to 16.67\%, showing a more substantial improvement. However, both still lag behind \qwentwo-7B. This gap likely stems from their earlier release stages, when HarmonyOS resources were more limited and training data quality was lower, leading to comparatively weaker coding capabilities than later versions in the same family.

\textbf{\deepseek-7B.}
An interesting observation arises with the \deepseek-7B model: without fine-tuning, it fails to pass any test cases in the HarmonyOS benchmark (i.e., a pass@1 of 0). Even after fine-tuning with OSS-Instruct~\cite{wei2024magicoder} data, its pass@1 only reaches 0.93\%, indicating negligible improvement. However, when fine-tuned with \app-generated data, the pass@1 increases significantly to 9.26\%, representing a 8.33\% gain over OSS-Instruct~\cite{wei2024magicoder}.

\textbf{Deepseek-LLM-7B.}
For DeepSeek-LLM-7B, the pass@1 score is 0\% without fine-tuning and improves to 7.41\% after fine-tuning with data generated by \app. The underlying reasons are generally consistent with those observed for Qwen-7B and Qwen2.5-7B, namely its earlier release stage and the relatively limited availability and quality of HarmonyOS-related training resources at that time, which constrain its coding capability compared to more recent model iterations.

\textbf{\mistral-7B.}
For the \mistral-7B model, the pass@1 score remains identical (4.63\%) both without fine-tuning and after fine-tuning with OSS-Instruct~\cite{wei2024magicoder} data. We attribute this to the OSS-Instruct~\cite{wei2024magicoder} dataset lacking sufficient API-related information, which limits the model’s ability to correctly generate HarmonyOS programs. By contrast, \app-generated data provides more comprehensive API information, enabling more effective fine-tuning. Consequently, the pass@1 for \mistral-7B improves to 14.81\% with \app, demonstrating a clear advantage over OSS-Instruct~\cite{wei2024magicoder}.

\textbf{Comparison with the non-fine-tuned model.}
Although \deepseek-671B is the full-scale version, it achieves only 3.70\% pass@1, likely due to limited HarmonyOS-related pretraining data. In contrast, \deepseek-7B and DeepSeek-LLM-7B fine-tuned with \app significantly outperform it. Similarly, despite its larger scale and broader corpus, \qwenthree-32B is surpassed by fine-tuned \qwentwo-7B and Qwen2.5-7B, further validating our fine-tuning strategy. Among commercial models, GPT-4o achieves the best performance without fine-tuning (17.59\% pass@1), yet still falls short of Qwen2.5-Coder-7B. Overall, fine-tuning with \app enables smaller models to outperform much larger counterparts.

\finding{In low-resource settings, model performance does not increase monotonically with scale; appropriate fine-tuning can even enable smaller models to outperform larger ones. Our \app method effectively enhances model capabilities under such conditions.}

\subsection{RQ2: Impact of Data Scale on Model Performance}

We systematically investigate the impact of data scaling and the ratio of single-API to multi-API data on LLM fine-tuning by training with varying volumes and compositions of synthetic code. To minimize random variability, experiments are conducted on three distinct local models across multiple data scales. This study aims to elucidate the effects of data volume and proportion on the training performance of 7B-parameter LLMs.

\parabf{Design.}

\textbf{Data Volume Experiment:} To investigate the impact of data volume on model fine-tuning, we randomly sample datasets of sizes 1,000, 2,000, 3,000, 4,000, 5,000, 6,000, 7,000, and 8,000, while preserving the original ratio between single-API and multi-API data. This allows us to examine how different data volumes affect model performance under identical fine-tuning conditions.

\textbf{Data Ratio Experiment:} To investigate the influence of the ratio between single-API and multi-API data on model fine-tuning, we fix the volume of multi-API data while varying the quantity of single-API samples. Specifically, we set the single-API to multi-API ratios to 1:9, 2:8, 3:7, 4:6, 5:5, 6:4, 7:3, 8:2, and 9:1, using approximate sample counts where exact division is not possible. This allows us to systematically examine how different data compositions affect model performance.

\parabf{Results.}

\begin{figure}[htbp]
  \centering
  \includegraphics[width=1.0\columnwidth]{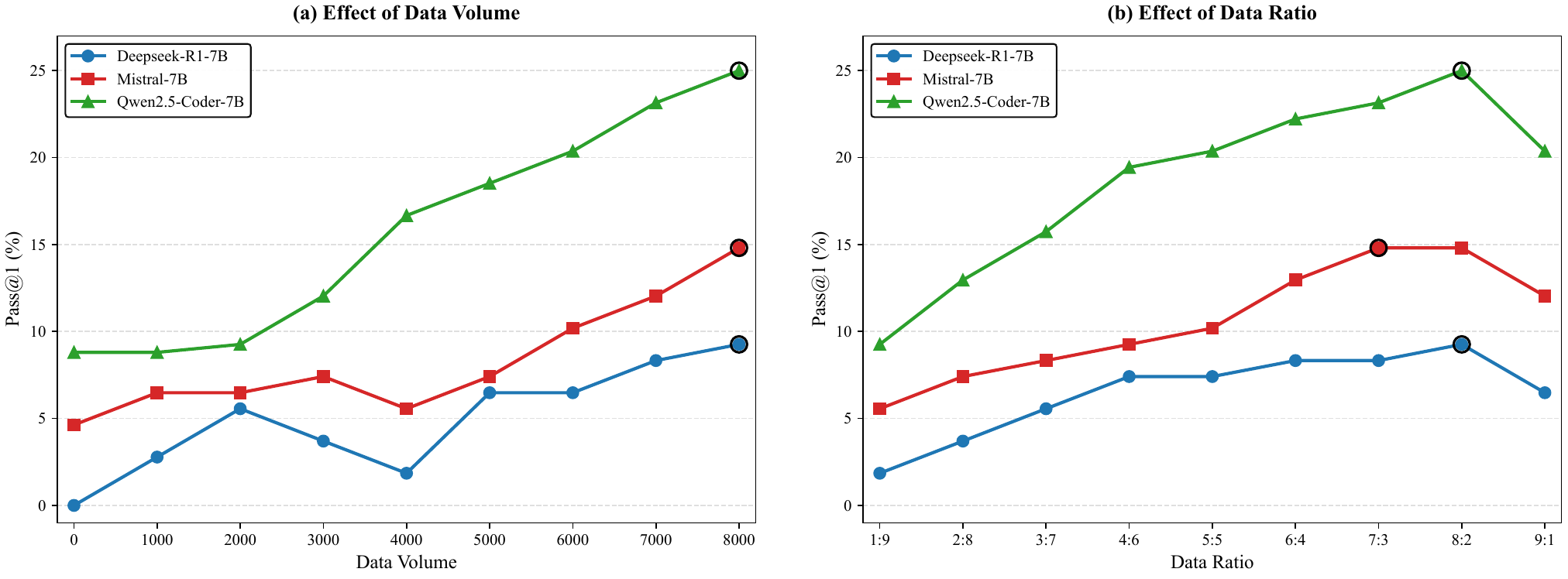}
  \caption{Impact of Data Scale and Data Ratio on Fine-tuning Performance.}
  \label{fig:data_scale}
\end{figure}

\textbf{Data Volume:} As illustrated in Figure~\ref{fig:data_scale} and Table~\ref{tab:data_volume}, \mistral-7B and \deepseek-7B exhibit a non-monotonic trend as fine-tuning data increases, a pattern consistent with the double descent phenomenon~\cite{doubledescent1} where generalization error temporarily rises near the interpolation threshold due to convergence toward sharp minima. However, as data volume continues to grow, stronger constraints steer the optimization toward flatter minima, thereby restoring performance. In sharp contrast, \qwentwo-7B bypasses this volatility; benefiting from robust pretrained representations and favorable initialization, it maintains stability in low-data regimes and achieves continuous improvement, suggesting that sufficiently expressive models can gain incrementally without undergoing transient degradation.

\textbf{Data Ratio:}
As shown in Figure~\ref{fig:data_scale} and Table~\ref{tab:data_ratio}, increasing the proportion of single-API to multi-API samples from 1:9 leads to a performance trend that first improves and then declines, with the 8:2 ratio achieving the best results under our setting.

This finding is consistent with prior work~\cite{codereasoningscaling, supervised}, which shows that combining a larger share of simpler examples with a smaller portion of complex ones can enhance code reasoning and supervised fine-tuning effectiveness. Accordingly, RQ1 adopts the empirically optimal 8:2 ratio.

\begin{table}[t]
\centering
\caption{The Impact of Data Ratio on Model Fine-Tuning.}
\label{tab:data_ratio}
\resizebox{\linewidth}{!}{
\begin{tabular}{lccccccccc}
\toprule
\textbf{Model} 
& 1:9 & 2:8 & 3:7 & 4:6 & 5:5 & 6:4 & 7:3 & 8:2 & 9:1 \\ 
\midrule

\deepseek-7B 
& 1.85\%  
& 3.70\%  
& 5.56\%  
& 7.41\%  
& 7.41\%  
& 8.33\%  
& 8.33\%  
& \textbf{9.26\%}
& 6.48\%  
\\

\mistral-7B 
& 5.56\%  
& 7.41\%  
& 8.33\%  
& 9.26\%  
& 10.19\% 
& 12.96\% 
& \textbf{14.81\%}
& \textbf{14.81\%}
& 12.04\%
\\

\qwentwo-7B 
& 9.26\%  
& 12.96\%  
& 15.74\%  
& 19.44\% 
& 20.37\%  
& 22.22\%  
& 23.15\%  
& \textbf{25.00\%} 
& 20.37\%
\\

\bottomrule
\end{tabular}
}
\end{table}
\begin{table}[t]
\centering
\caption{The Impact of Data Volume on Model Fine-Tuning.}
\label{tab:data_volume}
\resizebox{\linewidth}{!}{
\begin{tabular}{lccccccccc}
\toprule
\textbf{Model} 
& 0 & 1,000 & 2,000 & 3,000 & 4,000 & 5,000 & 6,000 & 7,000 & 8,000 \\ 
\midrule

\deepseek-7B 
& 0\% 
& 2.78\% 
& 5.56\% 
& 3.70\% 
& 1.85\% 
& 6.48\% 
& 6.48\% 
& 8.33\% 
& \textbf{9.26\%} \\

\mistral-7B 
& 4.63\% 
& 6.48\% 
& 6.48\% 
& 7.41\% 
& 5.56\% 
& 7.41\% 
& 10.19\% 
& 12.04\% 
& \textbf{14.81\%} \\

\qwentwo-7B 
& 8.80\% 
& 8.80\% 
& 9.26\% 
& 12.04\% 
& 16.67\% 
& 18.52\% 
& 20.37\% 
& 23.15\% 
& \textbf{25.00\%} \\

\bottomrule
\end{tabular}
}
\end{table}

\finding{ For API-oriented problem–code tuples, larger datasets generally improve fine-tuning performance, and smaller models may exhibit double descent. Performance is highest when the single-API to multi-API ratio is approximately 4:1, suggesting that a training distribution dominated by simpler API interactions with some complex examples supports effective adaptation and generalization.}

\subsection{RQ3: Distinction and Roles of Single-API and Multi-API Data}  
\parabf{Design.} To better investigate the impact of single-API and multi-API data, we conducted both quantitative and qualitative experiments. For the quantitative study, all multi-API samples were removed from the original 8,000-sample dataset to assess their effect on fine-tuning performance. Additionally, all single-API samples were also excluded, allowing us to examine the impact of a dataset composed solely of multi-API samples. For the qualitative study, the 1,600 multi-API samples and 6,400 single-API samples were partitioned into three groups, each containing 100 randomly selected single-API and multi-API samples. The sampling process was strictly random to ensure unbiased representativeness.

\textbf{Investigating the Impact of Removing Multi-API Data on Model Fine-Tuning.}
We selected the best-performing model, \qwentwo-7B, and fine-tuned it on the dataset with multi-API data removed, in order to investigate the impact of multi-API data.

\textbf{Investigating the Impact of Removing Single-API Data on Model Fine-Tuning.} To ensure consistency, the same models used in the aforementioned experiments were employed.

\begin{figure}[htbp]
    \centering
    \includegraphics[width=0.9\textwidth]{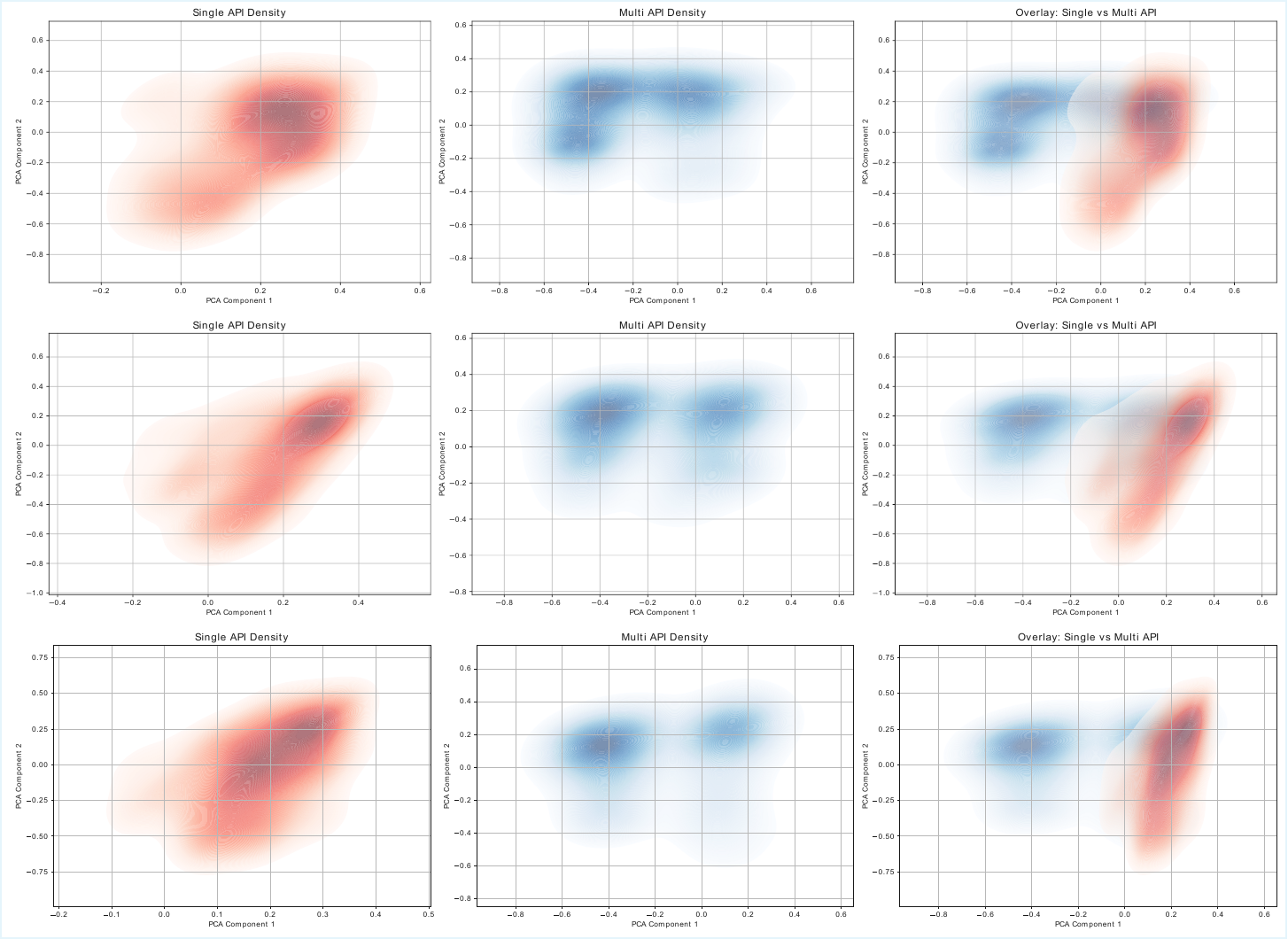}
    \caption{PCA visualizations for three groups (100 features each). Red/Blue points denote Single/Multi-API samples. The rightmost plot shows the combined distribution of both data types.}
    \label{fig:multi_single_API}
\end{figure}

\textbf{Investigating the Distribution of Different Types of API Data.}
To explore the relationship between multi-API and single-API data, we first approach it from a feature-based perspective. By embedding the questions and code in each sample, we obtain high-dimensional representations for both single-API and multi-API data. Principal Component Analysis (PCA) is then applied to reduce the dimensionality, resulting in the visualization shown in Figure~\ref{fig:multi_single_API}.

\parabf{Results.}
Table~\ref{tab:data_type} shows that removing multi-API data reduces the pass@1 score from 25.00\% to 10.19\%, indicating a substantial performance drop. This highlights the critical role of multi-API data in fine-tuning. Interestingly, when single-API data is removed, the performance decreases to 9.26\%, which is even lower than the 10.19\% observed when multi-API data is excluded. This underscores the fundamental importance of single-API data, which is further illustrated by the qualitative results below.
In the figure~\ref{fig:multi_single_API}, darker regions indicate areas where features are densely concentrated, while lighter regions correspond to sparser distributions. From the overlapping plots, it can be observed that multi-API data complements regions underrepresented by single-API data, as areas not covered by red are covered by blue. These observations provide qualitative evidence that multi-API data helps to fill the gaps left by single-API data

\begin{table}[t]
\centering
\caption{Impact of different training data types on fine-tuning performance.}
\label{tab:data_type}
\begin{tabular}{lccc}
\toprule
\textbf{Model} & \textbf{Single-API (6,400)} & \textbf{Multi-API (1,600)} & \textbf{All (8,000)} \\
\midrule
\qwentwo-7B & 10.19\% & 9.26\% & \textbf{25.00\%} \\
\bottomrule
\end{tabular}
\end{table}




\finding{Multi-API data can, to some extent, compensate for the knowledge deficiencies inherent in single-API data.}

\subsection{RQ4: Component Effectiveness}

\parabf{Design.}
To evaluate the contribution of each component in \app, we replace the MCTS retrieval strategy with Random and Frequency-based alternatives, and substitute the Knowledge Graph with plain text.

\textbf{MCTS Strategy Replacement.}
We assess MCTS by replacing it with: (1) a Random strategy that selects non-module nodes from the KG to construct multi-API data, and (2) a Frequency-based strategy that prioritizes APIs with higher in-degree and out-degree, reflecting stronger parameter or return-value connections.

\textbf{RAG-Based Direct Code Generation Without Fine-Tuning.}
Since MCTS relies on graph structures, we replace the knowledge graph with the original textual corpus and apply RAG on OHBen without fine-tuning, directly retrieving information to generate code for each query, thereby evaluating the effectiveness of structured knowledge compared to unstructured text.

\begin{table}[htbp]
  \centering
  \caption{Impact of different component strategies on model performance.}
  \label{tab:rq4}
  \begin{tabular}{l c c c}
    \toprule
    \textbf{Method} & \textbf{Search Strategy} & \textbf{Fine-tuning} & \textbf{Pass@1 (\%)} \\
    \midrule
    Plain Text + RAG & None & \ding{55} & 9.26 \\
    \midrule
    \multirow{2}{*}{\app (Variants)} & Random & \ding{51} & 20.37 \\
     & Frequency & \ding{51} & 22.22 \\
    \midrule
    \textbf{\app} & \textbf{MCTS} & \textbf{\ding{51}} & \textbf{25.00} \\
    \bottomrule
  \end{tabular}
\end{table}

\parabf{Results.}
As shown in Table~\ref{tab:rq4}, replacing the Knowledge Graph with a plain-text RAG approach achieves only 9.26\% Pass@1, slightly above direct generation (8.33\%), indicating limited benefit. This is mainly due to the large, definition-heavy source corpus, which hinders accurate retrieval for complex multi-API scenarios. Replacing MCTS with random and frequency-based strategies results in Pass@1 scores of 20.37\% and 22.22\%, respectively, both substantially lower than the full \app method. The random strategy often combines irrelevant APIs, while the frequency-based strategy captures parameter-coupled APIs but misses semantically related combinations without explicit parameter links. These results demonstrate that the LLM-derived UE value better captures scenario-level API similarity, enabling \app to construct higher-quality multi-API training data.

\finding{In data-scarce settings, RAG cannot replace synthetic data fine-tuning. Furthermore, UE-guided MCTS significantly outperforms baselines in generating API combinations, validating \app's effectiveness.}
\section{Threats to Validity}
\textbf{Internal Validity.}
A key threat to internal validity is the potential overlap between the training data and the benchmark, which could inflate model performance in RQ2 experiments on different data volumes. To address this, we used a fuzzy text-matching deduplication process (described in Section 2.3) to remove similar samples and maintain internal validity.

\textbf{External Validity.}
A major threat to external validity concerns the potential influence of model pretraining on fine-tuning, as models released later may have greater exposure to low-resource data over time. However, RQ1 findings indicate that although \deepseek-7B and \qwentwo-7B were released around the same time, \qwentwo-7B achieves better performance. Similarly, \mistral-7B, released earlier than \deepseek-7B, also demonstrates superior performance. These observations indirectly suggest that the timing and data of model pretraining have limited impact in low-resource scenarios.

\textbf{Threats to Generalizability.}
Our experiments and data generation were conducted only in low-resource scenarios on HarmonyOS, which may limit the generalizability of our findings. To test the broader applicability of the \app approach, we plan to extend our deployment to other low-resource environments in the future to validate the generalizability of our method.
\section{Related Work}

\subsection{Data Synthesis for Code LLMs}

In recent years, LLMs have shown remarkable capabilities in software engineering tasks such as code generation~\cite{brown2020languagemodelsfewshotlearners,feng-etal-2020-codebert,chen2021evaluatinglargelanguagemodels,ma2023bridgingcodesemanticllms,10.1145/3672456}, code completion~\cite{yu2025synthcodersyntheticalstrategytune,10.5555/3618408.3618894}, fault localization and repair~\cite{yang2023largelanguagemodelstestfree,samir2025improvedirbasedbuglocalization,xia2024agentlessdemystifyingllmbasedsoftware}, and code summarization~\cite{wu2024largelanguagemodelsserve,su2024contextawarecodesummarygeneration}. The performance of modern LLMs is largely dependent on the quality of the training data. Given that manually constructing high-quality and large-scale instruction datasets is both time-consuming and costly, leveraging LLMs to automatically generate synthetic data has become a key enabling technology driving progress in this field. 

Recent work explores synthesizing training data for code generation to improve LLMs without extensive manual annotation. Ma et al.~\cite{ma2025unitcoderscalableiterativecode} propose UnitCoder, a scalable pipeline that generates verifiable programs guided by unit tests, improving code LLMs via synthetic training data. Majumdar et al.~\cite{majumdar-etal-2025-genetic} introduce Genetic Instruct, an evolutionary method to scale instruction-tuning data for code generation by bootstrapping LLMs from seed samples. Shao et al.~\cite{shao-etal-2025-case2code} present Case2Code, a method to synthesize input-output behavior-based code generation data, enabling effective model training without human annotations. Gupta et al.~\cite{gupta2024targen} propose TarGEN, a self-correcting synthetic data generation framework using multi-step prompting for diverse NLP tasks, including code.

Existing approaches to data synthesis often rely on simple code snippets, fixed templates, or single-path API usage examples. In contrast, our work introduces subgraph-based prompts that explicitly encode element-level invocation and dependency relationships, leading to more realistic and semantically coherent data generation. Moreover, the adoption of a multi-path sampling strategy significantly enhances the diversity of the synthesized data.

\subsection{Domain-Specific and Low-Resource Code Generation}

Recent advances in code generation have increasingly addressed challenges associated with domain-specific languages (DSLs) and low-resource programming languages, such as limited data availability, unique syntax, and semantic constraints. Such as Bridge-Coder~\cite{zhang2024adcenhancingfunctioncalling} and SPEAC~\cite{mora2024syntheticprogrammingelicitationtexttocode}, enhance code generation for low-resource and DSLs by generating synthetic bridging data or intermediate representations that improve alignment between natural language and target code syntax. These techniques help mitigate data scarcity and improve syntactic and semantic correctness. Giagnorio et al.~\cite{giagnorio2025enhancingcodegenerationlowresource} further investigate fine-tuning strategies across different model sizes, showing that careful pretraining and adaptation significantly impact performance on languages like R and Racket. DreamCoder~\cite{ellis2020dreamcodergrowinggeneralizableinterpretable} takes a program synthesis approach, using a wake-sleep learning cycle to construct and expand DSLs, enabling the generation of structured, domain-relevant programs from natural language. AceCoder~\cite{li2023acecoderutilizingexistingcode} explores prompt engineering and retrieval-based augmentation by incorporating in-context examples selected from a code corpus, leading to substantial improvements in generation accuracy.

Substantial progress has been made in the ArkTS domain, with studies advancing tasks like code translation and analysis. Liu et al.~\cite{Liu2024LLMJavaArkTS} propose a Java-to-ArkTS converter that uses LLMs and the ThreadBridge library for shared-memory concurrency, enabling the migration of concurrent Java programs. Chen et al.~\cite{chen2025arkanalyzerstaticanalysisframework} introduce ArkAnalyzer, the first static analysis framework for ArkTS, capable of detecting performance bottlenecks and vulnerabilities. However, research on code generation for low-resource frameworks like ArkTS is still limited. Zhou et al.~\cite{porting} use an expert-curated knowledge base to translate TS/JS into ArkTS, while Liu et al.~\cite{liullm} apply template-constrained prompting for Java-to-ArkTS conversion. These methods often miss API usage modeling and lead to redundant re-implementations instead of reusing ArkTS APIs. In contrast, our approach generates high-quality training data from standard API documentation, with multi-API compositions and complex relationships, addressing data scarcity in low-resource scenarios. Unlike rigid prompt-based methods, our solution adapts to any API documentation without manual input, making it applicable to various API ecosystems.

\subsection{Knowledge Graph in Software Engineering}
Knowledge Graphs are widely used in software engineering to improve efficiency. In software maintenance, they help identify code dependencies and issues, making debugging and refactoring easier~\cite{8330232,xu2025astraautonomousspatialtemporalredteaming}. In code analysis, they provide a structured view of software components, assisting with static and dynamic analysis~\cite{Schiese2024,krishnan2025incrementalanalysislegacyapplications}. Knowledge Graphs are also used in automation, allowing intelligent systems to reason about data relationships, supporting tasks like code generation, repair~\cite{chen2025prometheusunifiedknowledgegraphs}, and testing. These applications highlight the potential of Knowledge Graphs in improving the software engineering life cycle.

In the domain of software engineering, researchers have established knowledge graphs for diverse objectives, encompassing domain terminology~\cite{fse2019glossary,wang2023xcos}, programming cases~\cite{fse22taskkg}, bug tracking~\cite{wang2017construct,su2021reducing}, API concepts~\cite{fse19apisummary,jos2021automatic,icsme2018apicaveat,ase20apicomp}, API documentation~\cite{icsme18docgen,icsme20docgen}. The knowledge graph we have constructed focuses on the HarmonyOS API documentation, with an emphasis on capturing the hierarchical and semantic relationships between different APIs (such as functional descriptions, input parameters, and return values). By structuring the real-world connections between APIs, the graph brings the representation closer to actual application scenarios, thus providing more realistic and practical support for generating training data for both single-API and multi-API use cases.
\section{Conclusion}
This paper introduces a method to generate API-oriented question-code datasets for low-resource scenarios using API knowledge graphs. Previous studies show that while LLMs are strong in reasoning about code logic, they often struggle with the syntax and API usage of low-resource languages. Our approach, \app, doesn’t require executable code and generates high-quality datasets purely from API documentation. This method is practical for companies or proprietary frameworks that want to fine-tune private LLMs without large-scale training data. Our experiment with HarmonyOS shows that a smaller model, \qwentwo, fine-tuned with API data, outperforms larger models like \gpt without fine-tuning, highlighting the benefits of API-oriented data for improving API knowledge in low-resource settings.

\section{Data Availability}
\label{sec:open-science}
To facilitate the replication study, we have released our data and code at :~\url{https://github.com/SYSUSELab/APIKG4SYN}.

\section*{Acknowledgments}
This work is supported by National Natural Science Foundation of China (No. 92582117, No. 62402113, No. 92582202), CCF-Huawei Populus Grove Fund CCF-HuaweiSE2025005, and GMCC-SYSU Joint Lab for Smart Applications.

\normalem

\bibliographystyle{ACM-Reference-Format}
\balance
\bibliography{refs}

\end{document}